\pdfoutput=1
\documentclass[conference]{IEEEtran}

\usepackage{lineno}
\modulolinenumbers[5]
\usepackage{nameref}

\usepackage{algorithm}
\usepackage[noend]{algpseudocode}
\usepackage{esvect}
\usepackage{tcolorbox}
\usepackage{csvsimple}
\usepackage{numprint}
\usepackage{xstring}
\usepackage{siunitx}

\newif\iftodo			  
\newif\ifchecklengths     
\newif\ifalternatives 	  

\todotrue
\checklengthsfalse
\alternativesfalse

\newlength{\aninch}
\usepackage{soul}
\usepackage{amssymb}
\usepackage{amsmath}
\usepackage[utf8]{inputenc}
\usepackage{graphicx}
\usepackage{epstopdf}
\usepackage{nth}
\usepackage{caption}
\usepackage{float}
\usepackage{multicol}
\usepackage[printonlyused,withpage]{acronym} 
\usepackage{mathtools}
\usepackage{color}
\usepackage{enumitem}
\usepackage{footnote}
\usepackage{url}
\usepackage{cleveref}
\usepackage{subcaption}
\usepackage{ulem}
\usepackage{tikz}
\usepackage{pgfplots}
\pgfplotsset{compat=1.8}
\usepackage{varwidth}
\usetikzlibrary{plotmarks}
\usetikzlibrary{backgrounds}
\usetikzlibrary{fit}
\usetikzlibrary{graphs}
\usepgfplotslibrary{groupplots}
\usepgfplotslibrary{colorbrewer}	
\usepgfplotslibrary{statistics}

\usetikzlibrary{chains,shapes.multipart}
\usetikzlibrary{shapes,calc,fit,arrows}
\usetikzlibrary{automata, positioning}

\usetikzlibrary{decorations.pathreplacing,decorations.markings,shapes.geometric,arrows,shapes,automata,backgrounds,petri}

\usetikzlibrary{graphs}

\makeatletter
\def\algbackskip{\hskip-\ALG@thistlm}
\makeatother

\makesavenoteenv{tabular}
\makesavenoteenv{table}

\usepackage[font={small}]{caption}

\usepgfplotslibrary{external}
\graphicspath{
	{figures/},
	{images/}
}
\usepgfplotslibrary{patchplots}
\pgfplotsset{select range/.style 2 args={
		x filter/.code={
			\ifnum\coordindex<#1\fi
			\ifnum\coordindex>#2\fi
		}
}}

\newacro{5G}{Fifth Generation Wireless Specifications}
\newacro{AP}{Application Providers}
\newacro{API}{Application Program Interface}
\newacro{AR}{Augmented Reality}
\newacro{ARQ}{Automatic Repeat Query}
\newacro{AWS}{Amazon Web Services}
\newacro{BER}{Benefit Effort Ratio}
\newacro{BER}{Bit Error Rate}
\newacro{CAPEX}{Capital Expenditure}
\newacro{CDN}{Content Delivery Network}
\newacro{CLI}{Command Line Interface}
\newacro{CPS}{Cyber-Physical System}
\newacro{CPU}{Central Processing Unit}
\newacro{CRC}{Cyclic Redundancy Check}
\newacro{CSI}{Channel State Information}
\newacro{DB}{Database}
\newacro{DC}{Data Center}
\newacro{DoF}{Degrees of Freedom}
\newacro{DOF}{Degrees Of Freedom}
\newacro{EC2}{Elastic Compute Cloud}
\newacro{FaaS}{Function-as-a-Service}
\newacro{FEC}{Forward Error Correction}
\newacro{FPGA}{Field-Programmable Gate Array}
\newacro{GUI}{Graphical User Interface}
\newacro{HARQ}{Hybrid Automatic Repeat Query}
\newacro{HW}{Hardware}
\newacro{IaaS}{Infrastructure as a Service}
\newacro{i.i.d}{Independent and Identically Distributed random variables}
\newacro{IoE}{Internet of Everything}
\newacro{I/O}{Input/Output}
\newacro{IoT}{Internet of Things}
\newacro{IP}{Infrastructure Providers}
\newacro{LDPC}{Low-Density Parity-Check }
\newacro{LTE}{Long Term Evolution}
\newacro{LuMaMi}{Lund Massive MIMO}
\newacro{MAC}{Medium Access Control}
\newacro{MAN}{Metropolitan Area Network}
\newacro{MCN}{Heterogeneous Distributed Computing}
\newacro{MCN}{Mobile Cloud Network}
\newacro{MD}{Mobile Device}
\newacro{MD}{Mobile Devices}
\newacro{MEC}{Mobile Edge Cloud}
\newacro{MIMO}{Multiple Input Multiple Output}
\newacro{MIP}{Mixed Integer Programming}
\newacro{mMTC}{massive Machine Type Communication}
\newacro{MN}{Mobile Network}
\newacro{MNO}{Mobile Network Operator}
\newacro{MNO}{Mobile Network Operators}
\newacro{MPC}{Model Predictive Control}
\acrodefplural{MPC}[MPCs]{Model Predictive Controllers}
\newacro{MPCer}[MPC]{Model Predictive Controller}
\newacro{MQTT}{Message Queue Telemetry Transport}
\newacro{MR}{Maximum-Ration Combining}
\newacro{MT}{Mobile Terminal}
\newacro{MU-MIMO}{Multi-User MIMO}
\newacro{NFV}{Network Function Virtualisation}
\newacro{NoOps}{No Operations}
\newacro{OFDM}{Orthogonal Frequency-Division Multiplexing}
\newacro{OPEX}{Operational Expenditure}
\newacro{PaaS}{Platform as a Service}
\newacro{PDC}{Proximal Data Centers}
\newacro{PID}{Proportional Integral Derivative}
\newacro{PM}{Physical Machine}
\newacro{QoS}{Quality of Service}
\newacro{QPSK}{Quadrature Phase Shift Keying}
\newacro{RAN}{Radio Access Network}
\newacro{RAT}{Radio Access Technology}
\newacro{RBS}{Radio Base Station}
\newacro{RBS}{Radio Base Stations}
\newacro{RDC}{Remote Data Centers}
\newacro{RTD}{Round-Trip Delay time}
\newacro{RTT}{Round Trip Time}
\newacro{RTT}{Round-Trip Time}
\newacro{SaaS}{Software-as-a-Service}
\newacro{SDK}{Software Development Kit}
\newacro{SDN}{Software Defined Networks}
\newacro{SDR}{Software Defined Radio}
\newacro{SLA}{Service Level Agreement}
\newacro{SLO}{Service Level Objective}
\newacro{SLO}{Service Level Objectives}
\newacro{SNR}{Signal-to-Interference-plus-Noise Ratio}
\newacro{SNS}{Simple Notification Service}
\newacro{SoS}{System of Systems}
\newacro{SP}{Service Providers}
\newacro{SQL}{Structured Query Language}
\newacro{SUMO}{Simulation of Urban MObility}
\newacro{SW}{Software}
\newacro{TLS}{Transport Layer Security}
\newacro{TraCI}{Traffic Control Interface}
\newacro{TSC}{Traffic Signal Control}
\newacro{TSP}{Transit Signal Priority}
\newacro{TTI}{Transmission Time Interval}
\newacro{UE}{User Equipment}
\newacro{UM}{User Mobility}
\newacro{URLLC}{Ultra-Reliable and Low-Latency Communication}
\newacro{UX}{User Experience}
\newacro{WAN}{Wide Area Network}
\newacro{WLAN}{Wireless Local Area Network}
\newacro{VM}{Virtual Machine}
\newacro{WSN}{Wireless Sensor Network}
\newacro{vSoftPLC}{virtual Software Programmable Logic Controllers}
\newacro{ZF}{Zero-Forcing}
\newacro{MS}{Mobile Station}
\newacro{NTP}{Network Time Protocol}
\newacro{PTP}{Precision Time Protocol}
\newacro{NGCC}{Next Generation Cloud Computing}
\newacro{ERDC}{Ericsson Research Data Center}
\newacro{DNR}{Distributed-NodeRED}
\newacro{ADC}{Analog to Digital Converter}
\newacro{DAC}{Digital to Analog Converter}
\newacro{NoOps}{No-Operations}
\newacro{PaaS}{Platform-as-a-Service}
\newacro{WASP}{Wallenberg Autonoms Systems and Software Program}
\newacro{LQR}{Linear Quadratic Regulator}
\newacro{VNF}{Virtualized Network Functions}
\newacro{CotC}{Control over the Cloud}
\newacro{ETSI}{European Telecommunications Standards Institute}
\newacro{AR}{Augmented Reality}
\newacro{AI}{Artificial Intelligence}
\newacro{ICT}{Information and Communication Technology}
\newacro{NCS}{Networked Control Systems}
\newacro{PLC}{Programmable Logic Controller}

\tikzset{naming/.style={align=center,font=\small}}
\tikzset{antenna/.style={insert path={-- coordinate (ant#1) ++(0,0.25) -- +(135:0.25) + (0,0) -- +(45:0.25)}}}
\tikzset{station/.style={naming,draw,shape=dart,shape border rotate=90, minimum width=10mm, minimum height=10mm,outer sep=0pt,inner sep=3pt}}
\tikzset{mobile/.style={naming,draw,shape=rectangle,minimum width=12mm,minimum height=6mm, outer sep=0pt,inner sep=3pt}}
\tikzset{radiation/.style={{decorate,decoration={expanding waves,angle=90,segment length=4pt}}}}

\tikzset{
reservoiri/.pic={
  \draw[line width=1pt]
    (0,0.25) --++ (0.5, -0.25) -- ++(0.5,0) -- ++(0,-0.5) -- ++(-0.5,0) -- ++(-0.5,-0.25);
   \node[above] at (0.5, 0.5) [text width=3cm, align=center] {#1};
   \coordinate (-input) at (0,-0.25);  
   \coordinate (-output) at (1,-0.25);  
  },
queuei/.pic={
  \draw[line width=1pt]
    (0,0) -- ++(1,0) -- ++(0,-0.5) -- ++(-1,0);
   \foreach \Val in {1,...,4}
     \draw ([xshift=-\Val*5pt]1,0) -- ++(0,-0.5);
   \node[above] at (0.5,0) {#1}; 
   \coordinate (-input) at (0,-0.25);  
   \coordinate (-output) at (1,-0.25);  
  },
rbs/.pic={
  \node[station] (base) {};

  \draw[line join=bevel] (base.100) -- (base.80) -- (base.110) -- (base.70) -- (base.north west) -- (base.north east);
  \draw[line join=bevel] (base.100) -- (base.70) (base.110) -- (base.north east);

  \draw[line cap=rect] ([xshift=-.1768cm,yshift=.6pt]base.north -| base.right tail) [antenna=1];
  \draw[line cap=rect] ([yshift=.6pt]ant1 |- base.north) -- node[above,shape=rectangle,inner ysep=+.3333em]{\dots} ([xshift=.1768cm,yshift=.6pt]base.north -| base.left tail) [antenna=2];

  \draw[thick,radiation,decoration={angle=45}] (.5,1.25) -- +(45:0.5);
  \draw[thick,radiation,decoration={angle=45}] (-.5,1.25) -- +(-45:-0.5);

  },
block/.pic = {
  \draw [line width=1pt] (0,0) rectangle (0.5,-0.5) node[pos=.5] {#1};
  \coordinate (-input) at (0,-0.25);  
  \coordinate (-output) at (0.5,-0.25); 
  }
}
\tikzset{naming/.style={align=center,font=\small}}
\tikzset{antenna/.style={insert path={-- coordinate (ant#1) ++(0,0.25) -- +(135:0.25) + (0,0) -- +(45:0.25)}}}
\tikzset{station/.style={naming,draw,shape=dart,shape border rotate=90, minimum width=10mm, minimum height=10mm,outer sep=0pt,inner sep=3pt}}
\tikzset{mobile/.style={naming,draw,shape=rectangle,minimum width=12mm,minimum height=6mm, outer sep=0pt,inner sep=3pt}}
\tikzset{radiation/.style={{decorate,decoration={expanding waves,angle=90,segment length=4pt}}}}

\usepackage{amsthm}
\theoremstyle{definition}

\newif\ifdisposition	  
\newif\ifobsdisposition   
\newif\ifremoved          
\newif\iftodo			  
\newif\ifunmasked			  

\todofalse
\dispositionfalse          
\obsdispositionfalse      
\removedfalse 		  
\unmaskedtrue

\usepackage{comment}
\ifremoved
\newenvironment{removed}{%
	\par\color{red}%
	\begin{small}\hrulefill\ \textit{Removed content} \hrulefill \end{small}\par%
}{%
	\par\begin{small}\hrulefill\ \textit{End removed} \hrulefill \end{small}\par%
}
\else
\excludecomment{removed}
\fi

\definecolor{aboutcolor}{rgb}{0.1,0.4,0.1}
\newcommand\about[1]{\ifdisposition \textcolor{aboutcolor}{ \begin{small}\textit{[About: #1]}\end{small}}\par \fi }

\ifunmasked
\else
\fi

\definecolor{kecolor}{rgb}{0,0,0.6}
\definecolor{pscolor}{rgb}{0.5,0.3,0.2}
\definecolor{jecolor}{rgb}{0.8,0.3,0.2}
\definecolor{mkcolor}{rgb}{0.8,0.3,0.2}
\definecolor{commentcolor}{rgb}{0,0,1}

\newcommand{\editorke}{\textcolor{kecolor}{\textbf{Karl-Erik: }}}
\newcommand{\editorps}{\textcolor{pscolor}{\textbf{Per: }}}

\newcommand{\editormk}{\textcolor{mkcolor}{\textbf{Maria: }}}

\newcommand\karlerik[1]{\iftodo \begin{small}\editorke \textcolor{commentcolor}{#1}\end{small} \fi}

\usepackage{suffix}
\newcommand\psnote[1]{\iftodo \begin{small}\editorps \textcolor{commentcolor}{#1}\end{small} \fi }
\WithSuffix\newcommand\psnote*[2]{\iftodo #2 \textcolor{commentcolor}{\begin{small}($\leftarrow$ \editorps #1)\end{small}} \else #2 \fi}
\newcommand\psfatal[1]{\iftodo \textcolor{commentcolor}{\begin{small}[\editorps #1]\end{small}} \fi}
\WithSuffix\newcommand\psfatal*[2]{\iftodo \textit{#2} \psfatal{#1} \else #2 \fi}
\newcommand\maria[1]{\iftodo \begin{small}\editormk \textcolor{commentcolor}{#1}\end{small} \fi}

\AtBeginDocument{\colorlet{defaultcolor}{.}}
\definecolor{calblue}{rgb}{0.337,0.478,0.51}
\definecolor{calgreen}{rgb}{0.471,0.733,0.259}
\definecolor{calorange}{rgb}{0.914,0.51,0.173}
\definecolor{calred}{rgb}{0.855,0.196,0.165}
\definecolor{darkgreen}{rgb}{0,0.6,0}

\crefname{appsec}{Appendix}{Appendices}

\begin{document}
	
	\title{\huge{An assisting Model Predictive Controller\\ approach to Control over the Cloud}
	}
	
	\ifunmasked
	\author{\IEEEauthorblockN{Per Skarin\IEEEauthorrefmark{1}\IEEEauthorrefmark{2},  Johan Eker\IEEEauthorrefmark{1}\IEEEauthorrefmark{2},  Maria Kihl\IEEEauthorrefmark{3}, Karl-Erik Årz{\'e}n\IEEEauthorrefmark{1}}
	\IEEEauthorblockA{\IEEEauthorrefmark{1}Department of Automatic Control, Lund University, Sweden}
	\IEEEauthorblockA{\IEEEauthorrefmark{2}Ericsson Research, Lund, Sweden}
	\IEEEauthorblockA{\IEEEauthorrefmark{3}Department of Electrical and Information Technology, Lund University, Sweden}}
	\fi
	
	\maketitle
	
	\begin{abstract}
In this paper we develop a computational offloading strategy with graceful degradation for executing Model Predictive Control using the cloud. Backed up by previous work we simulate the control of a cyber-physical-system at high frequency and illustrate how the system can be improved using the edge while keeping the computational burden low.
	\end{abstract}
	
	
	\section{Introduction}
\ifchecklengths
\begin{tabular}{ll}
	textwidth & \the\textwidth\\
	textheight & \the\textheight\\
	hoffset & \the\hoffset\\
	voffset & \the\voffset\\
	marginparwidth & \the\marginparwidth\\
	marginparpush & \the\marginparpush\\
	footskip & \the\footskip\\
	inch & \the\aninch
\end{tabular}\\
\fi

\about{Add mobile edge clouds as a motivator}
Edge computing \cite{etr-2018-11-5g-and-distributed-cloud,1808.05283,etsi_wp11_mec_a_key_technology_towards_5g} is a means to provide highly responsive and reliable cloud services through local data centers, possibly closely integrated with 5G \ac{RAN}. By providing compute and storage in close proximity to users, edge computing achieves low latency and high reliability. Compared to large data centers, a typical edge node may only serve a limited amount of applications, but in return provide low communication overhead and reduce the load on the core networks. One incarnation of this is mobile base station virtualization for use with \ac{NFV}, which is an industry standard for virtual network functions~\cite{wp-5g-systems}. In addition to host network functions, such as routing and access management, 5G edge nodes may utilize the same virtualized infrastructure to also support user applications. This provides the mentioned benefit of low latency, but also provides the potential for end user context awareness (e.g. location, speed, etc.) and collaboration, without involving back-end services. Third party applications could operate at the base stations in concert with the underlying telecommunication infrastructure and further leverage the ultra reliable low latency communication (URLLC) capabilities of 5G~\cite{wp-5g}.
These characteristics provide an interesting environment for  control of critical systems.


\ifunmasked
In previous work~\cite{8473376} we examined this setup and its potential for automatic control in combination with 5G.
\fi
A modern edge cloud and \ac{IoT} research test-bed was 
developed but the control application remained traditional. The test-bed hardware contained a local device connected to the physical process under control, a compute server directly connected to a 5G base station representing a local edge data center, and two remote data centers. A \ac{MPCer}\acused{MPC} was used to control a physical ball and beam process. 
The support for software migration provided by the platform~\cite{Persson2015} was used to randomly migrate the \ac{MPC} calculations between the different hardware nodes, including the local device, creating varying delays and jitter for the control loop. In~\cite{8473376}, however, these delays were largely ignored, i.e., no delay compensation techniques were applied. Also, it was assumed that the local device was able to execute the compute-intensive \ac{MPC} controller, something which is not assumed here.
\ifunmasked
The test-bed still serves as a conceptual basis for this work, but we now introduce cloud nativeness to the domain of automatic control and what we refer to as Control over the Cloud. 
\fi
We consider cloud applications for critical time-sensitive cyber-physical systems and present an approach from the domain of automatic control.
The approach shows how cloud strategies can improve control systems while keeping some of the formal guarantees, e.g., stability.

The illusion of infinite compute and storage resources that the cloud and the edge/fog provides opens up a number of interesting possibilities for control applications. The resources can be used for executing more advanced control strategies, e.g., based on online optimization and learning using massive data sets, than what is possible on the local device. 
The cloud can scale resources with the problem and implement efficient strategies for each computation. This allows the controller to evaluate complex problems which are too computationally demanding to perform locally.
Information made available through the communication network, e.g., additional more complex models and information about other similar application types, can be incorporated and used to improve the control, avoiding the overhead and potential concerns of communicating this information to the local system. The drawback of moving online computations into the cloud is the additional delays that it creates. However, by placing some of the computations at the edge this can also be managed.

\begin{figure}[h]
	\ifunmasked
	\includegraphics[width=\linewidth]{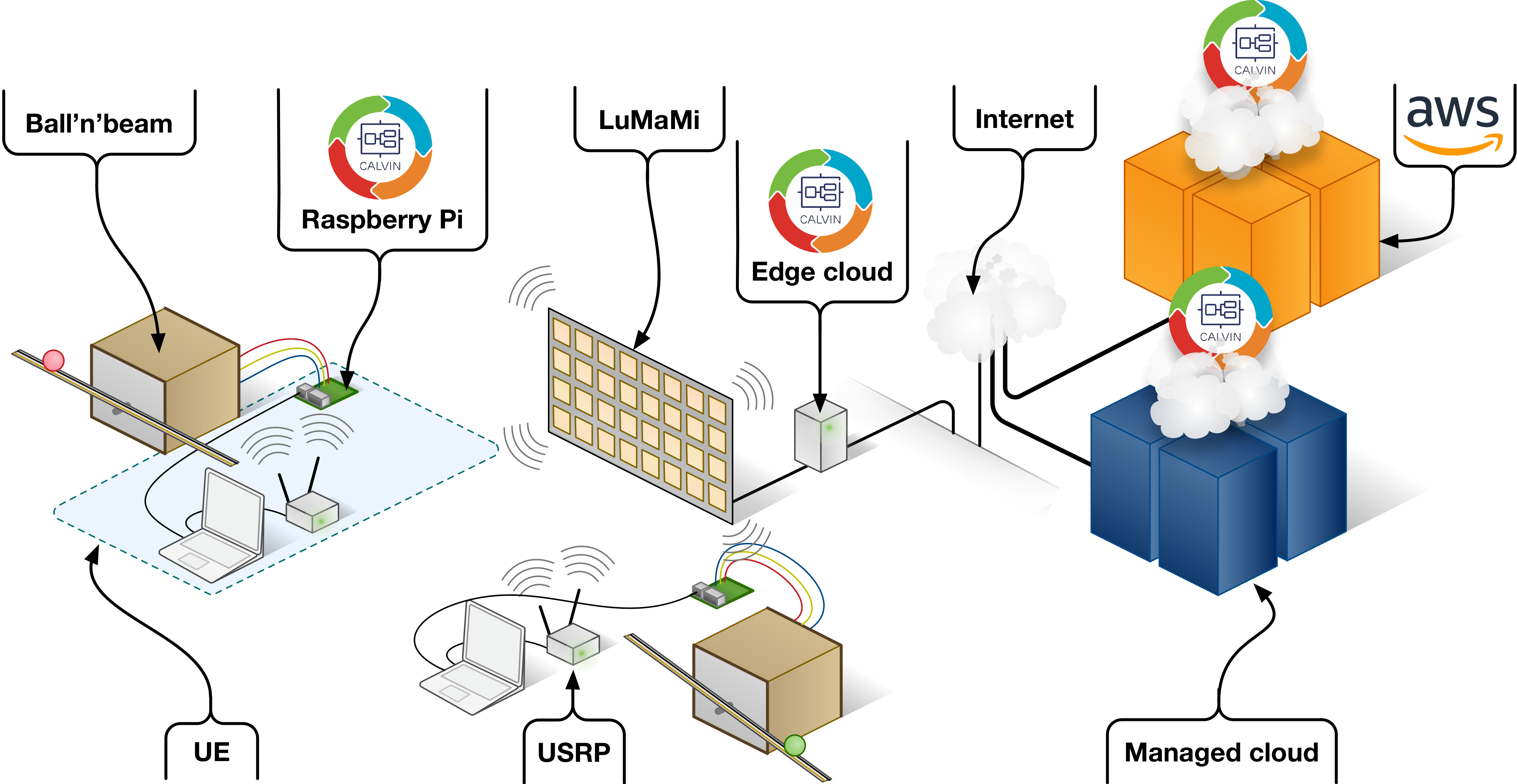}
	\caption{Fog/Edge cloud test bed operating a model predictive controller on an IoT and cloud service platform with seamless software migration support}	  
	\fi
	\label{fig:testbed}
\end{figure}

\about{Paper outline}
Here a cloud-assisted control approach is presented that uses the cloud to execute a number of model-based optimizations, each using different optimization parameters. When cloud connectivity is available the proposed approach improves upon the control performance obtained by the controller executing in the local device.
\Cref{sec:problem} presents the problem and \Cref{sec:relatedwork} gives an overview of related work.
\Cref{sec:feedbackcontrol} introduces the basic control theory for multiple-input-multiple-output (MIMO) linear systems necessary to build the formal case for our cloud-assisted control approach.
\Cref{sec:assistedcontrol} provides the structure for the controller aimed at edge clouds and \Cref{sec:evaluation} evaluates the approach using simulation. Finally,
\Cref{sec:conclusions} summarizes and concludes the paper.

	\section{Problem description}\label{sec:problem}
The purpose of this paper is to develop a control architecture that can operate reliably and seamlessly using the cloud.
In the cloud, execution times and network latency can exhibit large random variations in the short term while over long time the cloud system itself also evolves. This makes the cloud environment highly stochastic and chaotic. A cloud native control design takes this into account while making use of on demand resources to improve its operation.

The system should use the abundance of resources provided by the cloud but should also be able to scale down, or gracefully degrade, when the cloud is not available.
A basic requirement in automatic control is closed loop system stability.
That is, for a bounded input the output is also bounded.
Stability must be ensured in the transition to and from the degraded mode of operation.
The goals of the cloud control system can be summarized as:
\begin{enumerate}
	\item The system should provide acceptable control performance and stability also in case of connectivity loss.
	\item Under ordinary conditions when the cloud is available performance, reliability, quality and/or safety of the system should be improved.
\end{enumerate}
This work also emphasizes a design with a single controller objective, a single model and a single set of controller constraints, rather than, e.g., switching between different controllers or using a hierarchy of controllers, each defined using different objectives.
The controller should operate efficiently at frequencies in the range of a few to a hundred Hertz. 
All applications in the cloud must consider tail latencies~\cite{dean2013tail} and potential connectivity loss but this frequency range is of particularly interest for industrial automation~\cite{Hegazy2015,07396554,07879156}. \psnote{K-E, godkänner du detta?}

	\section{Related work} 
\label{sec:relatedwork}
Control-over-networks is a branch of \ac{NCS} which has been studied extensively focusing on aspects of network delay, packet dropout, channel capacity and security. A survey of research in the domain of \ac{NCS} is provided in \cite{Xia2015} including control-over-networks and \ac{MPC}.
In addition to the topics covered by traditional \ac{NCS} the cloud brings in a new aspect in terms of on-demand resources, also commonly referred to as elastic compute.
Another networked control consideration is that of centralized versus distributed methods~\cite{Richards2007,Rawlings2009,camponogara2002distributed,Christofides2013}.
With fast interconnects, computational power and adjacency to large amounts of collected data, cloud data centers are good candidates for the execution of distributed methods to improve efficiency through horizontal scaling.
Individual optimizations performed in the cloud, as proposed in this paper, can very well be implemented using a combination of sequential, parallel and singleton strategies which can be assessed, scaled and deployed at runtime.

It is not uncommon that control of cyber-physical-systems is considered in terms of replacing existing systems with virtualized equivalents.
One example is the case study in~\cite{6837587} where hardware \acp{PLC} are replaced by software counterparts in terms of virtual machines in a private cloud.
In~\cite{7879156} offloading controllers to the cloud is considered for industrial \ac{IoT}.
The paper studies the effects of delays, mitigations techniques at the device and an adaptive proportional-integral controller.
Similar to~\cite{8473376} the primary intention of these papers is to showcase and benchmark existing technologies.
Hegazy and Hefeeda introduce industrial automation as a service in~\cite{Hegazy2015}.
The proposed service provides delay mitigation through the introduction of artificial delays, an adaptive Smith Predictor, and moving average and moving variance delay estimates.
A fault tolerance scheme is suggested using multiple Internet links and cloud providers.
The service is exemplified through \ac{PID} control and sampling periods of 200 ms for fault tolerance and 300 ms for the delay compensation.
The proposed use cases include redundant controller for critical-systems and temporary control-over-the-cloud during maintenance.

Our work in control-over-the-cloud aims to advance research with new approaches to controller design, creating elastic control structures and autonomous resiliency.

\section{MPC and LQR control}\label{sec:feedbackcontrol}	

\about{This brief paragraph is intended to keep the readers interest (the cloud stuff is coming soon) while some necessary background is established.}
In this section we will introduce a strategy for control of cyber-physical-systems that utilize the cloud.
The strategy combines linear \ac{MPC} with the \acf{LQR}.
The following provides a background to the method through an overview of these two controllers and their relation.

\newcommand{\captionfiglqr}{A figure which examplifies what can happen when LQR goes out of linearized model region/saturates input. Perhaps combined with MPC in the same figure.}
\newcommand{\captiontrajectories}{A figure which examplifies what can happen when LQR goes out of linearized model region/saturates input. Perhaps combined with MPC in the same figure.}

\subsection{Model predictive control}\label{sec:mpc}
\about{A straight forward run-through on MPC. Sets the keywords optimization, model, constraints and horizon. Intentional oversight on the part of terminal cost and state as it will be covered later.}
\acfp{MPC} use on-line numerical optimization to calculate the control signal to apply as input to the system under control, commonly referred to as \textit{the plant} or \textit{the process}.
A discrete-time linear \ac{MPC} is specified by \Cref{eq:lqmpc}.
It uses a quadratic cost function $l(x_k, u_k) = x^T_k Q x_k + u^T_k R u_k$, a cost $P$ applied to the final state (refered to as the \textit{terminal cost}), a system model defined by matrices $A$ and $B$, and inequality and equality constraints set by the matrix vector pairs $G,g$ and $H,h$ respectively.
The cost matrix $Q$ penalizes moving away from the desired state while $R$ penalizes the control signal. 

\begin{equation}\label{eq:lqmpc}
\begin{split}
\underset{\bf{u}}{\text{minimize}}\> J =\> &\sum_{i=k}^{k+N-1} x_i^{T}Qx_i + u_i^{T}Ru_i + x_{k+N}^T P x_{k+N} \\
\text{subject to}\quad &x_{i+1} = Ax_i + Bu_i\\
&G \begin{bmatrix} x_i\\u_i \end{bmatrix} \leq g,\> H \begin{bmatrix} x_i\\u_i \end{bmatrix} = h
\end{split}
\end{equation}


Each time the plant is sampled the controller must first use the controller input and the controller output to calculate the state, $x_k$, to be used as input to the optimization. This is done using a state estimator or \textit{observer}.
It then translates the  states and constraints based on the  \textit{set-point}, e.g., instead of using the actual state the error between the actual state and desired states is used.
This information is fed into the optimization routine which uses the model to predict and optimize the plant behavior and generates a sequence of input signals that will drive the system state towards the set-point.
To be robust to uncertainty and disturbances only the first input signal in the control signal sequence is used and the whole process is repeated again at the next sampling instant.

How far into the future the controller predicts the system state trajectory and the length of the calculated control signal sequence are determined by the \textit{horizon}.
This is given by the parameter $N$ in \Cref{eq:lqmpc}.
If the horizon is too short the optimization problem may become infeasible, i.e., no control signal can be found that fulfills the constraints, which might lead to instability.
A large horizon is computationally expensive and may take too long time to evaluate.
In practice one therefore attempts to limit the horizon and/or pre-compute the solution to the optimization problem. The latter, known as \textit{explicit \ac{MPC}}, is outside the scope of this paper.

The major benefit of \ac{MPC} compared to other controller types is the possibility to enforce constraints on, e.g., the process state, output and/or control signal. 
The number of iterations in the optimizer needed to solve the problem varies depending on whether the constraints are active or not, e.g., if the state is close to a constraint or not, which may cause large execution time variations. This can be caused by a disturbance acting on the system or by an unfortunate set-point selection causing the system to come too close to the constraints.

In the proposed cloud-assisted controller, \ac{MPC} is used at the cloud level, either in an edge data center or in a remote data center. In order to solve the problem of which prediction horizon to chose, at each sample a number of \ac{MPC} requests are sent to the cloud, each with a different horizon. These are then executed in parallel, possibly at different nodes. High-priority requests could be executed in an edge data center to increase the probability that the results of these requests are returned in time, whereas requests that are deemed less important could be executed in a remote data center.

\subsection{Linear quadratic regulator}\label{sec:lqr}
\about{Quick overview of LQR, then go into MPC complications and dual mode.}
The \acf{LQR} is a special case of \Cref{eq:lqmpc}.
We arrive at it by removing all constraints and setting the horizon ($N$) to infinity, resulting in \Cref{eq:lqr}.
\begin{equation}\label{eq:lqr}
\begin{split}
\underset{\bf{u}}{\text{minimize}}\> J = & \sum_{k=0}^{\infty} x_{k}^{T}Qx_{k} + u_k^{T}Ru_k\\
\text{subject to} \quad & x_{k+1} = Ax_{k} + Bu_k
\end{split}
\end{equation}
This form has an analytical solution~\cite{kalmanlqr} which provides a globally optimal state feedback controller.
That is, given a model expressed by $A, B$ and cost matrices $Q, R$ it is possible to compute a gain matrix $K$ that when multiplied with the state vector $x_k$, generates a control law $u_k = -Kx_k$ which is an optimal controller given the specification in Equation (\ref{eq:lqr}).
To follow a set-point, $x_k$ is replaced by a state error defined as $x_e =  x_{sp} - x_k$ and the control signal is created by applying the \textit{control law} $u_k = -Kx_e$.


The \ac{LQR} provides a globally optimal controller, can handle \ac{MIMO} systems, provides good stability and gain margins, is fast to execute, and easy to define and implement~\cite{kalmanlqr}.
However, it cannot handle constraints, i.e. it is assumed to operate in an infinite state and control signal space.
As a result, the controller may require very large control signals and/or force the system into unwanted and unrecoverable states.
To avoid this, designers must impose restrictions such as limiting the operating range of the controller (restrict the possible set-points) or tweak the costs in the optimization problem. These restrictions make the \ac{MPC} a more attractive solution.

In the proposed cloud-assisted controller the LQR will be the basis for the controller used at the local device level.

\subsection{\ac{MPC} stability and dual mode control}\label{sec:mpcpractice}
\about{Motivate the LQR for stability}
\ac{LQR} control  is stable and always has a well defined control action. 
The \ac{MPC} in \Cref{eq:lqmpc} is, however, not guaranteed to be stable, and could fail to return a control action if the optimization becomes infeasible.
A method for achieving stability in the \ac{MPC} is to combine it with an \ac{LQR}.
Since  \Cref{eq:lqr} is a special case of \Cref{eq:lqmpc} it is straightforward to generate a \ac{LQR} from the \ac{MPC} specification.
The method that ensures stability and feasibility with a limited horizon is to require that the final state of the optimization, $x_{k+N}$, falls into a set of states for which \ac{LQR} is guaranteed to fulfill the constraints.
The \ac{MPC} approximates the control actions of the \ac{LQR} when the system moves away from the constraints and approaches the set-point.
For a nominal system and fixed set-point, it can be shown that if a limited horizon controller with such a requirement finds a feasible solution it will be stable and remain feasible. 
We refer to the limited set of final states as the terminal set, $\mathcal{T}$ and the formal requirement is that $x_{k+N} \in \mathcal{T}$.
Whether this requirement holds depends on the length of the horizon, $N$, and the initial state $x_0$.	
If the terminal set is introduced formally into the optimization then $N$ must be large enough to reach $\mathcal{T}$ for all possible $x_0$, otherwise the optimization becomes infeasible and does not provide a control action.
It may be hard to determine how large $N$ must be and it is often the case that $\mathcal{T}$ is reached although $N$ is not large enough to ensure this will happen.
The requirement to reach $\mathcal{T}$ is therefore often replaced by extensive testing, which can allow for a smaller $N$.
In this manner $N$ is a compromise between computation time, the operating range of the controller, and reliability. 

\about{Now state why this is important to our case}
The take away from this, considering implementing \ac{MPC} using the cloud, is that the stabilized \ac{MPC} is composed of an \ac{LQR} which guarantees stability and feasibility but which has a limited operating range, and a constrained \ac{MPC} optimization which extends the operating range of the controller.
The state space in which the controller works depends on the horizon.

	\section{Assisted feedback control}\label{sec:assistedcontrol}
The execution time of an \ac{MPC} varies and can be considerable~\cite{8473376}.
It is especially problematic when operating close to constraints which is also when a good control response is most critical.
A method to mitigate the execution delay is to introduce an artificial constant delay of one sample that is accounted for in the controller
\ifunmasked
~\cite{Cortes2012,Arzen_4}.
\fi
Rather than using the current state $x_k $ to calculate and apply $u_k(0)$ under the assumption that no time has passed from sensory readings to control action, instead one predicts $x_{k+1}$ and then one has the entire sampling period available to calculate the vector $\vv{u}_{k+1}$. At the start of the next sample period $u_{k+1}(0)$ is applied.
In the cloud-assisted feedback controller, a control delay of one sample is introduced and the obtained time interval is used to execute an assisted controller using the cloud.
The approach can also be extended to time delays longer than one sample.

\subsection{Assisted mode}\label{sec:assistedmode}
In the assisted mode the controller is connected to the network and continuously receives control actions from the \ac{MPC} (\Cref{sec:mpc}) in the cloud.
With every sample a new request is sent to the cloud service.
This request includes updated state information and the horizons to evaluate. Several horizons are evaluated to increase the chances of receiving a response in time. Results may be lost or late due to network delay, computation time, admission time into the cloud services, packet loss, connectivity loss and machine failure. Short horizons may not provide feasible solutions and long horizons can take too long to evaluate. 
Methods for deciding which range of horizons to use is not considered in this work.

\psnote{Här nedan har jag utökat med lite argument för kort horisont}
At the start of the next iteration the local controller selects one of the available responses.
Since all the responses returned in time guarantee stability and feasibility the result from the shortest horizon can be selected and used.
This can be useful in several ways.
Subsequent requests may choose to use shorter horizons to reduce the computational burden and cost of operating in the cloud.
Short horizons may be routed to a less powerful but more reliable system such as an edge node or run on the local device.
Short horizons must also reach the terminal set in fewer steps, this quickly puts the system in a good state for the conservative local mode to take over in case of connection issues.
If no response was received the controller uses a combination of control signals from previous MPC evaluations and its \ac{LQR} controller during a transition period after which only the local controller is used.

\begin{removed} 
	Response times are a combination of service admission time, network delays and execution time.
	Specialized edge clouds reduce the network delays and could improve admission time.
	This provides additional execution time for the evaluation.
	On the other hand, capacity in the edges can limit the number of consecutive  evaluations and the efficiency of optimizations.
\end{removed}

\begin{figure}
	\resizebox{\columnwidth}{!}{\input{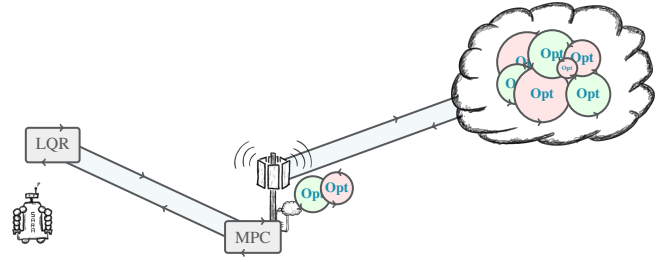}}
	\caption{The assisted controller runs an \ac{LQR} as a backup controller that continuously receives support from MPC optimizations executing in the cloud.}
	\label{fig:assistedrobot}
\end{figure}

\subsection{Local mode}
In local mode the control is achieved using only \ac{LQR} (\Cref{sec:lqr}).
Local mode is entered when connectivity is lost.
It is also possible to define conditions under which local mode gives sufficiently good performance and therefore can be used without cloud support also if the latter is available.
This could, e.g., be for cost or energy saving reasons.
To operate reliably set-point changes can be restricted in the local mode.
Beyond these restrictions assisted mode is necessary for optimal performance.

\subsection{Switching from local to assisted mode}
The switch from local to assisted mode is instantaneous.
Whenever the controller receives one or more responses in time from the cloud it will apply one of those as the next control output.
Similarly, if the controller is in the process of moving from the assisted to the local mode, that process is immediately interrupted and the control output is replaced.

\subsection{Switching from assisted to local mode}
The switch from assisted to local mode is critical.
If local mode is entered immediately when connectivity is lost or individual results are delayed the system can suffer large constraint violations. This can be seen in the example in Section V.E. The switching strategy employed here makes use of the fact that when the \ac{MPC} is evaluated it not only returns the control signal to be applied to the process but also a sequence of future control signals based on the predicted trajectory of the system. Under ideal circumstances, e.g., without disturbances, the system can operate in {\it open loop} using these control signals, although they have been calculated based on old information.

If the predictions correspond well with the actual evolution of the system these control signals can simply be applied until the sequence is exhausted.
This requires a high precision system model, correct actuation (low noise), no external disturbances and that the set-point does not change during the transition.
It is required of the assisted mode \ac{MPC} that its last action puts the system in a state which can be handled by the local mode \ac{LQR}. Hence, it would under ideal conditions be possible to execute in open loop during the duration of the control sequence and thereafter switch to local \ac{LQR} mode. However, to accommodate systems which require closed-loop control,  a strategy is used which instead is based on a weighted control law that gradually shifts the dominant control signal from the one obtained from the cloud using \ac{MPC} to the one obtained using local \ac{LQR} control, according to  below,
\begin{equation}
\begin{split}
u_k = \beta(k) \kappa_{l}(\cdot) + (1-\beta(k)) \kappa_{r}(\cdot),\ \beta \in [0, 1]\\
\end{split}
\label{eq:gradual}
\end{equation}
where $\kappa_{l}(\cdot)$ and $\kappa_{r}(\cdot)$ are the local and remote control laws.
In the local mode $\beta$ is set to one.
When data arrives from the network the immediate response is to prioritize the remote control law through a small $\beta$.
If no further data arrives the subsequent inputs are read from the available open loop sequence while $\beta$ is increased to put more emphasis on the local controller.
However, it should be noted that this approach does still not guarantee that the constraints are not violated.

The effect of the gradual shift is shown by the example in Section V.E,
Not using the open loop control signals causes the \ac{LQR} to move far away from the constraint region. Using open loop control signals only can cause large deviations from the optimal path due to disturbances and  model errors.
The gradual shift in the strategy also ensures a smooth transition from assisted to local mode.

\subsection{Example}\label{sec:example}
\maria{Hur har du gjort detta? Ganska svårt att förstå för en icke-reglertekniker, behövs en beskrivning på hur du fått fram resultaten}
The following section provides an example to illustrate the operation of the assisted controller.
The example consists of a regulator problem where the objective of the controller is to force the system state to the origin.
The second order model of the system under control and the cost matrices are shown in \Cref{eq:exmodel} and \Cref{eq:excost}.
\begin{equation}
A = \begin{bmatrix}0.9752 & 1.4544 \\ -0.0327 & 0.9315 \end{bmatrix} \quad B = \begin{bmatrix} 0.0248 & 0.0327 \end{bmatrix}^T
\label{eq:exmodel}
\end{equation}
\begin{equation}
Q = \begin{bmatrix} 10 & 0 \\ 0 & 10 \end{bmatrix} \quad R = 1
\label{eq:excost}
\end{equation}
Using this we arrive at the \ac{LQR} gain matrix in \Cref{eq:exlqr}
\begin{equation}
K = \begin{bmatrix} 1.6478 & 11.8344 \end{bmatrix}
\label{eq:exlqr}
\end{equation}
The \ac{MPC} is further defined by the constraint specification in \Cref{eq:exconstraints}.
For details on how to setup and operate the \ac{MPC} we refer to the literature~\cite{Rawlings2009,BorrelliBemporadMorari}.
\begin{equation}
\begin{aligned}
G &= \begin{bmatrix} 1 & 0 \\ 0 & 1 \\ -1 & 0 \\ 0 & -1 \\ 1 & -1 \end{bmatrix}, g = \begin{bmatrix} 5 \\ 0.2 \\  5 \\  0.2 \\ 1.75 \\ 1.75 \end{bmatrix}\\
\end{aligned}
\label{eq:exconstraints}
\end{equation}

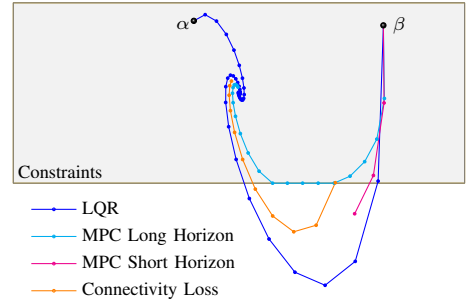
\begin{figure}[h]
	\centering
\begin{tikzpicture}
\begin{axis}[
	legend style={font=\scriptsize, fill opacity=0, text opacity=1, anchor=south west, draw=none, at={(0.1, 0.01)}},
	ylabel={$x_1$},
	xlabel={$x_2$},
	axis lines=none,
	legend cell align=left,
	x=0.6cm,
	y=6cm
]
\addplot[patch, gray!10, patch type=rectangle] coordinates {(-5,-0.2) (5,-0.2) (5,0.2) (-5,0.2)};
\addplot[patch, mesh, gray, patch type=rectangle] coordinates {(-5,-0.2) (5,-0.2) (5,0.2) (-5,0.2)};


	\addplot[mark=*, mark options={scale=0.25}, blue] table [col sep=comma, x index=1, y index=2] {data/seminar/pathfollow_modelerrorN10.csv};
	\addlegendentry{LQR}
	
	\addplot[mark=*, mark options={scale=0.25}, blue] table [col sep=comma, x index=1, y index=2] {data/seminar/pathfollow_clinv.csv};

	\addplot[mark=*, mark options={scale=0.25}, cyan] table [col sep=comma, x index=5, y index=6] {data/seminar/pathfollow_modelerrorN10.csv};
	\addlegendentry{MPC}

	\addplot[mark=*, select range={0}{3}, mark options={scale=0.25}, magenta] table [col sep=comma, x index=5, y index=6] {data/seminar/pathfollow_N5.csv};
	\addlegendentry{MPC short horizon}	

	\addplot[select range={5}{15}, mark=*, orange, mark options={scale=0.25}] table [col sep=comma, x index=9, y index=10] {data/seminar/pathfollow_modelerrorN15.csv};

	\addplot[mark=ball, mark options={scale=0.5, ball color=black}] coordinates {(-1,0.16)};
	\addplot[mark=ball, mark options={scale=0.5, ball color=black}] coordinates {(3.2,0.15)};	
	
	\legend{,,LQR,, MPC Long Horizon, MPC Short Horizon, Connectivity Loss}
	\pgfplotsset{
		after end axis/.code={
			\node[black,anchor=west] at (axis cs:-1.6,0.15){\scriptsize{$\alpha$}};			
			\node[black,anchor=west] at (axis cs:3.2,0.15){\scriptsize{$\beta$}};
			\node[black,anchor=south west] at (axis cs:-5.1,-0.21){\scriptsize{Constraints}};			
		}
	}
x\end{axis}
\end{tikzpicture}
\caption{Simulation of \ac{LQR} and \ac{MPC}}
\label{fig:lqr_and_mpc}
\end{figure}

\Cref{fig:lqr_and_mpc} shows the system state space and four different control trajectories for two starting positions $\alpha$ and $\beta$. The gray rectangle marks the state constraints.
When starting in $\alpha$ the system never reaches the constraints independently of which controller that is used, while when
starting from $\beta$ the \ac{LQR} (in blue) violates the constraints.
The short horizon \ac{MPC} (in red) does not initially see the constraints and starts similarly to the \ac{LQR}. However, later it becomes infeasible. The long horizon \ac{MPC} generates a control trajectory at the constraint boundary.
The last trajectory in orange shows what happens if the \ac{MPC} with the long horizon is disconnected after a few steps and the \ac{LQR} immediately takes over, i.e., without the weighting strategy.

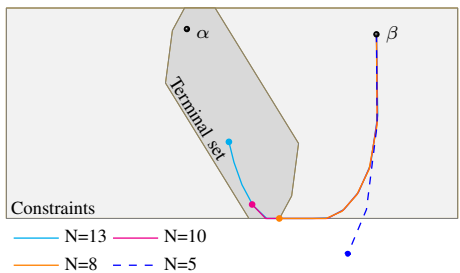
\begin{figure}
	\centering

\begin{tikzpicture}

\begin{axis}[
legend style={font=\scriptsize, fill opacity=0, text opacity=1, anchor=south west, draw=none, at={(0.08, -0.02)}, legend columns=2},
ylabel={$x_1$},
xlabel={$x_2$},
axis lines=none,
legend cell align=left,
x=0.6cm,
y=7cm
]

\addplot[patch, gray!10, patch type=rectangle] coordinates {(-5,-0.2) (5,-0.2) (5,0.2) (-5,0.2)};

\addplot[patch, mesh, gray, patch type=rectangle] coordinates {(-5,-0.2) (5,-0.2) (5,0.2) (-5,0.2)};

\addplot[patch, gray!30, line width=0.1pt, patch type=polygon, vertex count=8] coordinates {
	(   -1.4742,    0.0574)
	(   -1.3214,    0.1572)
	(   -1.0521,    0.2000)
	(   -0.3744,    0.2000)
	(   1.4742,   -0.0574)
	(   1.3214,   -0.1572)
	(   1.0521,   -0.2000)
	(   0.3744,   -0.2000)
};
\addplot[patch, mesh, gray, line width=0.1pt, patch type=polygon, vertex count=8] coordinates {
	(   -1.4742,    0.0574)
	(   -1.3214,    0.1572)
	(   -1.0521,    0.2000)
	(   -0.3744,    0.2000)
	(   1.4742,   -0.0574)
	(   1.3214,   -0.1572)
	(   1.0521,   -0.2000)
	(   0.3744,   -0.2000)
};


	\addplot[mark=*, mark indices={14}, line width=0.5, select range={0}{13}, mark options={scale=0.5}, cyan] table [col sep=comma, x index=5, y index=6] {data/seminar/pathfollow_N13.csv};

	\addplot[mark=*, mark indices={11}, line width=0.5, select range={0}{10}, mark options={scale=0.5}, magenta] table [col sep=comma, x index=5, y index=6] {data/seminar/pathfollow_N10.csv};
	
	\addplot[mark=*, mark indices={9}, line width=0.5, select range={0}{8}, mark options={scale=0.5}, orange] table [col sep=comma, x index=5, y index=6] {data/seminar/pathfollow_N8.csv};
	

	\addplot[mark=*, mark indices={4}, line width=0.5, select range={0}{3}, mark options={scale=0.5}, dashed, blue] table [col sep=comma, x index=5, y index=6] {data/seminar/pathfollow_N5.csv};

	\addplot[mark=ball, mark options={scale=0.5, ball color=black}] coordinates {(-1,0.16)};
	\addplot[mark=ball, mark options={scale=0.5, ball color=black}] coordinates {(3.2,0.15)};	

	\legend{,,,, N=13, N=10, N=8, N=5}
	
	\pgfplotsset{
		after end axis/.code={
			\node[black,anchor=west] at (axis cs:-1,0.15){\scriptsize{$\alpha$}};
			\node[black,anchor=west] at (axis cs:3.2,0.15){\scriptsize{$\beta$}};
			
			\node[black,anchor=south west] at (axis cs:-5.1,-0.21){\scriptsize{Constraints}};
			\node[black,anchor=south west, rotate=-58] at (axis cs: -1.7,0.065){\scriptsize{Terminal set}};
		}
	}
\end{axis}
\end{tikzpicture}
	\caption{Control trajectories of \acp{MPC} with different horizons.}
	\label{fig:cloudcontrol}
\end{figure}

\Cref{fig:cloudcontrol} shows four \ac{MPC} control trajectories, with different horizons, starting in $\beta$ and the region for which the \ac{LQR} always stays within the constraints.
The latter is known as the \textit{terminal set}.
Three things can be observed. First, it can be seen that $\alpha$ lies inside the terminal set and $\beta$ does not, as expected from the trajectories in \Cref{fig:lqr_and_mpc}.
Second, short horizons lead to infeasibility and constraint violation.
Third, the longest horizon is not  necessary since there are shorter alternatives that give the same trajectory.
This motivates why multiple MPC invocations with different horizons are requested.

\begin{figure}
	\centering
	\begin{tikzpicture}
\begin{axis}[
legend style={font=\scriptsize, fill opacity=0, text opacity=1, anchor=south west, draw=none, at={(0.08, 0.18)}, legend columns=2},
ylabel={$x_1$},
xlabel={$x_2$},
axis lines=none,
legend cell align=left,
x=0.6cm,
y=7cm
]

\addplot[patch, gray!10, patch type=rectangle] coordinates {(-5,-0.2) (5,-0.2) (5,0.2) (-5,0.2)};
\addplot[patch, mesh, gray, patch type=rectangle] coordinates {(-5,-0.2) (5,-0.2) (5,0.2) (-5,0.2)};

\addplot[patch, gray!30, line width=0.1pt, patch type=polygon, vertex count=8] coordinates {
	(   -1.4742,    0.0574)
	(   -1.3214,    0.1572)
	(   -1.0521,    0.2000)
	(   -0.3744,    0.2000)
	(   1.4742,   -0.0574)
	(   1.3214,   -0.1572)
	(   1.0521,   -0.2000)
	(   0.3744,   -0.2000)
};
\addplot[patch, mesh,, gray, line width=0.1pt, patch type=polygon, vertex count=8] coordinates {
	(   -1.4742,    0.0574)
	(   -1.3214,    0.1572)
	(   -1.0521,    0.2000)
	(   -0.3744,    0.2000)
	(   1.4742,   -0.0574)
	(   1.3214,   -0.1572)
	(   1.0521,   -0.2000)
	(   0.3744,   -0.2000)
};

\addplot[white] coordinates {(5.5, 0.25)};	
\addplot[white] coordinates {(5.5, -0.35)};

\addplot[mark=*, mark options={scale=0.25}, orange] table [col sep=comma, x index=7, y index=8] {data/seminar/pathfollow_modelerrorN10.csv};

\addplot[mark=*, line width=0.5, select range={0}{10}, mark options={scale=0.25}, cyan] table [col sep=comma, x index=5, y index=6] {data/seminar/pathfollow_N10.csv};

	\addplot[mark=ball, mark options={scale=0.5, ball color=black}] coordinates {(3.2,0.15)};	
	
	\pgfplotsset{
	after end axis/.code={
			\node[black,anchor=west] at (axis cs:3.2,0.15){\scriptsize{$\beta$}};
			
			\node[black,anchor=south west] at (axis cs:-5.1,-0.21){\scriptsize{Constraints}};
			\node[black,anchor=south west, rotate=-58] at (axis cs: -1.7,0.065){\scriptsize{Terminal set}};			
		}
	}
\legend{,,,,,,Dual mode in open loop, Predicted path}
\end{axis}
\end{tikzpicture}
	\caption{Applying open loop sequence on connection loss. The deviation is due to a disturbance caused by a model error.}
	\label{fig:openloop}
\end{figure}
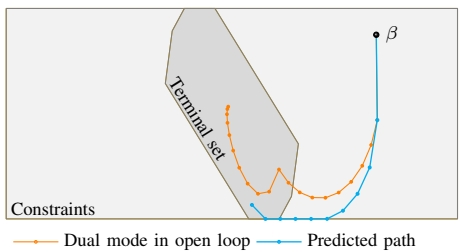

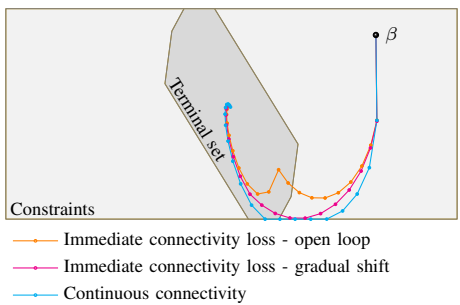
\begin{figure}
	\centering
\begin{tikzpicture}
\begin{axis}[
legend style={font=\scriptsize, fill opacity=0, text opacity=1, anchor=south west, draw=none, at={(0.08, 0.04)}, legend columns=1},
ylabel={$x_1$},
xlabel={$x_2$},
axis lines=none,
legend cell align=left,
y=7cm,
x=0.6cm
]

\addplot[patch, gray!10, patch type=rectangle] coordinates {(-5,-0.2) (5,-0.2) (5,0.2) (-5,0.2)};
\addplot[patch, mesh, gray, patch type=rectangle] coordinates {(-5,-0.2) (5,-0.2) (5,0.2) (-5,0.2)};

\addplot[patch, gray!30, line width=0.1pt, patch type=polygon, vertex count=8] coordinates {
	(   -1.4742,    0.0574)
	(   -1.3214,    0.1572)
	(   -1.0521,    0.2000)
	(   -0.3744,    0.2000)
	(   1.4742,   -0.0574)
	(   1.3214,   -0.1572)
	(   1.0521,   -0.2000)
	(   0.3744,   -0.2000)
};
\addplot[patch, mesh, gray, line width=0.1pt, patch type=polygon, vertex count=8] coordinates {
	(   -1.4742,    0.0574)
	(   -1.3214,    0.1572)
	(   -1.0521,    0.2000)
	(   -0.3744,    0.2000)
	(   1.4742,   -0.0574)
	(   1.3214,   -0.1572)
	(   1.0521,   -0.2000)
	(   0.3744,   -0.2000)
};
\addplot[white] coordinates {(5.5, 0.25)};	
\addplot[white] coordinates {(5.5, -0.35)};

\addplot[mark=*, mark options={scale=0.25}, orange] table [col sep=comma, x index=7, y index=8] {data/seminar/pathfollow_modelerrorN10.csv};

\addplot[mark=*, mark options={scale=0.25}, magenta] table [col sep=comma, x index=11, y index=12] {data/seminar/pathfollow_modelerrorN10.csv};

\addplot[mark=*, mark options={scale=0.25}, cyan] table [col sep=comma, x index=5, y index=6] {data/seminar/pathfollow_modelerrorN10.csv};



	\addplot[mark=ball, mark options={scale=0.5, ball color=black}] coordinates {(3.2,0.15)};

	\pgfplotsset{
	after end axis/.code={
			\node[black,anchor=west] at (axis cs:3.2,0.15){\scriptsize{$\beta$}};
			
			\node[black,anchor=south west] at (axis cs:-5.1,-0.21){\scriptsize{Constraints}};
			\node[black,anchor=south west, rotate=-58] at (axis cs: -1.7,0.065){\scriptsize{Terminal set}};			
		}
	}
\legend{,,,,,,Immediate connectivity loss - open loop, Immediate connectivity loss - gradual shift, Continuous connectivity}
\end{axis}
\end{tikzpicture}
\caption{The result of applying an exponential gradual decline to the open loop sequence in order to manage the model error disturbance.}
\label{fig:finale}
\end{figure}

In \Cref{fig:openloop}, $N=10$ has been selected but runs in open-loop (in orange).
Consider that immediately after receiving the result of the optimization starting in $\beta$ the cloud is disconnected.
All 10 input signals from the optimization are used before the \ac{LQR} takes over.
The deviation from the predicted optimal path (in blue) occurs because of a model error.
\Cref{eq:exmodelerr} shows the system model used in the simulation. 
Compared to the model in \Cref{eq:exmodel} there is an error of approximately 2.7\% in the first parameter.
The gradual shift in \Cref{eq:gradual} provides a simple and generic compromise between the \ac{LQR} mode which can compensate for disturbances and the open loop which takes constraints into account.
The effect is shown in \Cref{fig:finale}. Here, three trajectories are shown: running the \ac{MPC} in open loop with a small model error causing a deviation from the nominal trajectory (in orange), using the gradual shift in \Cref{eq:gradual} (in magenta), and finally the trajectory generated without any connectivity loss (in blue), i.e., the trajectory generated by the \ac{MPC} executing in closed loop. \karlerik{En avslutande mening som sammanfattande säger att gradual shift är bra, dvs upprepar det som man kan se i figuren.}

\begin{equation}
A = \begin{bmatrix}0.95 & 1.4544 \\ -0.0327 & 0.9315\end{bmatrix}
\label{eq:exmodelerr}
\end{equation}

\newcommand{\captionblockdiagram}{Block diagram of local and remote controllers}

	\section{Evaluation}\label{sec:evaluation}
\maria{Detta är väldigt mycket reglerteknik, vilket är lite synd för artikeln måste ha en större målgrupp än så, speciellt som den ska skickas till Edge. Intro och problembeskrivning är mer networking}
The approach is evaluated using a simulated ball and beam process.
The process consists of a tilting beam with a metal ball rolling along the beam.
The objective is to balance the ball at a certain set-point position.
The measured variables are the beam angle and the ball position.
The control signal is the voltage to the motor that tilts the beam.
The model for this consists of a triple integrator, i.e., a third-order linear system~\cite{Virseda2004}.
The sampling frequency is 20 Hz, i.e., the \ac{MPC} calculations must be returned within 50 ms. The simulation is implemented in Matlab using Simulink and TrueTime, a Simulink toolbox for simulating distributed real-time systems with real-time kernels and networks~\cite{cervin2003does}.
The \ac{MPC} is implemented using the Matlab \textit{quadprog} command.
\ifunmasked
The reason for not using the edge cloud test-bed from \cite{8473376} and a physical process is that here the focus is control performance which requires controlled and deterministic conditions to be evaluated properly.
\fi
For example, the performance of the cloud assisted controller and its reaction to networking conditions is evaluated using a nominal system without external disturbances and noise.
\maria{Detta räcker inte. Hur ser din simuleringsmodell ut? Som det är nu så går det inte att återskapa dina experiment}
A conservative \ac{LQR} is assisted using an \ac{MPC} with a limited terminal state.
The critical constraint is the beam's endpoints.
We want to be able to move the ball as close to the endpoints as possible without losing it. 

\Cref{fig:evaluation-control-plot}-I shows the execution of the \ac{LQR} as specified in \Cref{sec:lqr} and \Cref{eq:exlqr}.
The horizontal lines at 0.55 and -0.55 show the hard constraints at the two ends of the beam.
The red line shows the trajectory of the ball using the \ac{LQR} controller and a set-point signal (in dashed gray) with maximum and minimum values 0.52 and -0.52 respectively. 
This controller violates the constraints and will cause the ball to fall off the beam.
The blue line shows the same \ac{LQR} operated with a limited set-point range of $abs(x_{sp}-x_k) \leq 0.4$.
This gives smaller overshoots and the constraints are met.
The goal is now to use a cloud-assisted controller design with an \ac{MPC} that can approach the reaction time of the original \ac{LQR} without any constraint violations.

\newcommand\reflinecolor{black}
\newcommand\reflinewidth{1}
\newcommand\lqrinfcolor{red}
\newcommand\lqrlimcolor{blue}

\newcommand\largehcolor{green}
\newcommand\smallhcolor{red}

\newcommand\goodnetcolor{green}
\newcommand\badnetcolor{blue}
\pgfplotsset{every axis title/.append style={at={(0.5,0.9)}}}

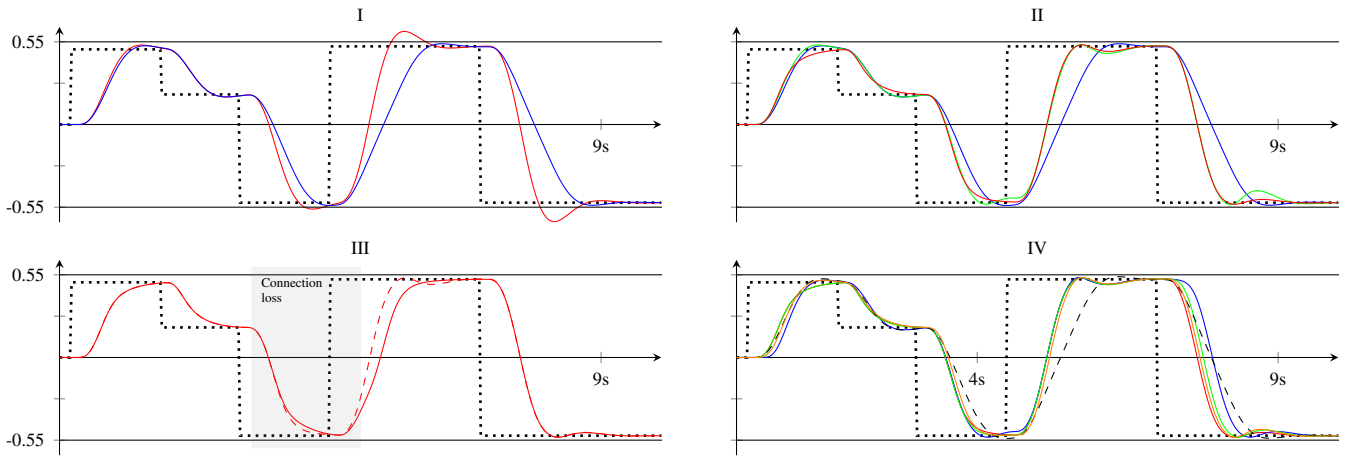
\begin{figure*}[t]
    \centering
	\begin{tikzpicture}
	\begin{groupplot}[
		group style={group size=2 by 2,vertical sep=0.5cm},
		xmin=0,xmax=20,
		legend style={font=\scriptsize},
		legend pos=south west,
		legend columns=-1,
		y=2cm,
		x=0.8cm
	]
	
	\nextgroupplot[mark=none, xmin=0, xmax=10,
	every x tick label/.append style={font=\scriptsize},
	every y tick label/.append style={font=\scriptsize},
	title=\scriptsize I,
	axis lines=center,
	ytick={-0.55, -0.275, 0, 0.275, 0.55},
	yticklabels={-0.55,,,,0.55},	
	xtick={9}, xticklabels={9s},
	ymin=-0.65, ymax=0.65
	]
	
	\addplot[\reflinecolor, dotted, line width=\reflinewidth] table[header=false, col sep=comma, x index=0, y index=1] {data/evaluation/complex_lqr_inf/ref.csv};
	
	\addplot[\lqrinfcolor] table[header=false, col sep=comma, x index=0, y index=1] {data/evaluation/complex_lqr_inf/plant.csv}; \label{pgfplots:ecp-lqrinf};
	
	\addplot[\lqrlimcolor] table[header=false, col sep=comma, x index=0, y index=1] {data/evaluation/complex_lqr_0_4/plant.csv}; \label{pgfplots:ecp-lqrlim};
	
	\addplot[black, domain=0:20] {0.55};
	\addplot[black, domain=0:20] {-0.55};
	
	\nextgroupplot[mark=none, xmin=0, xmax=10,
	every x tick label/.append style={font=\scriptsize},
	every y tick label/.append style={font=\scriptsize},
	title=\scriptsize II,
	axis lines=center,
	ytick={-0.55, -0.275, 0, 0.275, 0.55},
	yticklabels={,,,,},	
	xtick={9}, xticklabels={9s},
	ymin=-0.65, ymax=0.65
	]
	
	\addplot[\reflinecolor, dotted, line width=\reflinewidth] table[header=false, col sep=comma, x index=0, y index=1] {data/evaluation/complex_cloud_long_mpconly/ref.csv};
	
	\addplot[\lqrlimcolor] table[header=false, col sep=comma, x index=0, y index=1] {data/evaluation/complex_lqr_0_4/plant.csv};	
	
	\addplot[\largehcolor] table[header=false, col sep=comma, x index=0, y index=1] {data/evaluation/complex_cloud_long_mpconly/plant.csv};	\label{pgfplots:ecp-largeh}
	
	\addplot[\smallhcolor] table[header=false, col sep=comma, x index=0, y index=1] {data/evaluation/complex_cloud_short_mpconly/plant.csv}; \label{pgfplots:ecp-smallh}
	
	\addplot[black, domain=0:20] {0.55};
	\addplot[black, domain=0:20] {-0.55};

	\nextgroupplot[mark=none, xmin=0, xmax=10,
	every x tick label/.append style={font=\scriptsize},
	every y tick label/.append style={font=\scriptsize},
	title=\scriptsize III,
	axis lines=center,
	ytick={-0.55, -0.275, 0, 0.275, 0.55},
	yticklabels={-0.55,,,,0.55},	
	xtick={9}, xticklabels={9s},
	ymin=-0.65, ymax=0.65
	]
	
	\addplot[\reflinecolor, dotted, line width=\reflinewidth] table[header=false, col sep=comma, x index=0, y index=1] {data/evaluation/complex_cloud_short_netdown/ref.csv};
	
	\addplot[\smallhcolor] table[header=false, col sep=comma, x index=0, y index=1] {data/evaluation/complex_cloud_short_netdown/plant.csv}; \label{pgfplots:ecp-netdown}
	\addplot[\smallhcolor, dashed] table[header=false, col sep=comma, x index=0, y index=1] {data/evaluation/complex_cloud_short_mpconly/plant.csv}; \label{pgfplots:ecp-netdownref}
	
	\addplot[black, domain=0:20] {0.55};
	\addplot[black, domain=0:20] {-0.55};	
	
	\pgfplotsset{
		after end axis/.code={
			\draw[fill,gray,opacity=0.1] (axis cs:3.2,0.60) rectangle (axis cs:5,-0.60);
			\node[anchor=west] at (axis cs:3.2,0.50) {\tiny Connection};
			\node[anchor=west] at (axis cs:3.2,0.40) {\tiny loss};		
		}
	}

	\nextgroupplot[mark=none, xmin=0, xmax=10,
	every x tick label/.append style={font=\scriptsize},
	every y tick label/.append style={font=\scriptsize},
	title=\scriptsize IV,
	axis lines=center,
	ytick={-0.55, -0.275, 0, 0.275, 0.55},
	yticklabels={,,,,},	
	xtick={4, 9}, xticklabels={4s, 9s},
	ymin=-0.65, ymax=0.65
	]
	
	\addplot[\reflinecolor, dotted, line width=\reflinewidth] table[header=false, col sep=comma, x index=0, y index=1] {data/evaluation/lambda_like/ref.csv};
	
	\addplot[\smallhcolor] table[header=false, col sep=comma, x index=0, y index=1] {data/evaluation/complex_cloud_short_mpconly/plant.csv}; \label{pgfplots:ecp-lossref}
	
	
	\addplot[\badnetcolor] table[header=false, col sep=comma, x index=0, y index=1] {data/evaluation/complex_cloud_short_dropout_badnet/plant.csv}; \label{pgfplots:ecp-badnet}
	
	\addplot[\goodnetcolor] table[header=false, col sep=comma, x index=0, y index=1] {data/evaluation/lambdalike_upper/plant.csv}; \label{pgfplots:ecp-goodnet}	
	
	\addplot[black,dashed] table[header=false, col sep=comma, x index=0, y index=1] {data/evaluation/complex_lqr_0_4/plant.csv}; \label{pgfplots:ecp-lqrlim2};
	
	\addplot[orange] table[header=false, col sep=comma, x index=0, y index=1] {data/evaluation/withedge/plant.csv}; \label{pgfplots:ecp-edge};
	
	\addplot[black, domain=0:20] {0.55};
	\addplot[black, domain=0:20] {-0.55};

	\end{groupplot}
	\end{tikzpicture}
	\caption{I: LQR with full set-point range (\ref{pgfplots:ecp-lqrinf}) and limited set-point range (\ref{pgfplots:ecp-lqrlim}). The full range constraint violations will move the ball off the beam. II: Continously operating long horizon MPC (\ref{pgfplots:ecp-largeh}) and cloud assisted short horizon selection (\ref{pgfplots:ecp-smallh}) and the limited range LQR (\ref{pgfplots:ecp-lqrlim}). III: Degradation due to connection loss when using short range selection (\ref{pgfplots:ecp-netdown}). IV: Degradation due to latency distributions when using short range selection. Delay distributions are shown in \Cref{fig:evaluation-distributions-plot}. Orange (\ref{pgfplots:ecp-edge}) line is with an additional edge node using distribution B in \Cref{fig:evaluation-distributions-plot}. Red line (\ref{pgfplots:ecp-lossref}) is controller without network delay and the dashed line (\ref{pgfplots:ecp-lqrlim2}) is LQR with limited set-point.}
	\label{fig:evaluation-control-plot}	
\end{figure*}

\Cref{fig:evaluation-control-plot}-II shows two assisted controllers and compares them with the \ac{LQR} with the limited set-point range.
In every sample, the assisted controllers send requests to evaluate \ac{MPC} instances of horizons of length \numrange{6}{22}.  This range of horizons has been selected based on experience in controlling the ball and beam and to illustrate the effects of the assisted control.
The green line shows a controller which always selects the longest horizon available within the \SI{50}{\ms} time frame.
In this example, all feasible solutions are available in time and the green line is therefore roughly equivalent to operating a single controller using a horizon of \num{22}.
Roughly equivalent only, because it can be seen from \Cref{table:usage} that there are some occasions when no \ac{MPC} result at all is returned, which then results in a gradual switch to the \ac{LQR} controller.
The red line in \Cref{fig:evaluation-control-plot}-II instead always selects the shortest available horizon.
Due to the terminal state constraint $x_f  \in \mathcal{T}$ this results in a more conservative control with a gradually decreasing horizon after a set-point change.
For reasons outlined in \Cref{sec:assistedmode} this mode is chosen for further evaluation.


\Cref{fig:evaluation-control-plot}-III illustrates the effect of a 1.8 seconds connection loss, indicated by the time interval in gray.
The controller enters the gradual switching mode and eventually runs in pure \ac{LQR} mode until connectivity is restored (shown using a full red line). 
The dashed line  shows the assisted controller without network disconnect.
The limited range \ac{LQR} is not shown since it is clear that it represents the lower bound performance which would occur with a very long disconnect.

\Cref{fig:evaluation-control-plot}-IV, finally, shows instead what happens when the total delay not only consists of \ac{MPC} execution delays but also of network delays. 
Two scenarios are shown. The red line is again the result of the cloud assisted controller in \Cref{fig:evaluation-control-plot}-II (using the shortest available horizon).
The response time includes both \ac{MPC} execution time and the system (network/scheduling) delay.
The statistics of the \ac{MPC} execution time is shown in \Cref{fig:evaluation-exectimes-plot}.
These execution times are obtained using the number of iterations required for the Matlab \textit{quadprog} command to finish the optimization, scaled linearly with the horizon and multiplied by a benchmark time from a modern laptop.
It was determined that an horizon of \num{20} required \SI{1}{ms} and with the number of iterations given by $\gamma$ the execution time in seconds is given by \Cref{eq:exectime}.

\begin{equation}
\tau_{exec} = 0.001\cdot \gamma \cdot N/20
\label{eq:exectime}
\end{equation}

The number of iterations varies with the set-point and the state of the plant.
The execution times in \Cref{fig:evaluation-exectimes-plot} are added to the network delays given by the delay distribution plots in \Cref{fig:evaluation-distributions-plot} returning the total delays of the MPC invocations.
If the total delay sums up to more than \SI{50}{\ms} the results are dropped. 
The distribution in \Cref{fig:evaluation-distributions-plot}-A was chosen as the upper part of a bimodal distribution from measurements of requests to Amazon Lambda~\cite{tarnebergthesis}.\psnote{Om vi bara använder det gröna så fyll i med det här} 
The distribution in \Cref{fig:evaluation-distributions-plot}-B covers a larger part of the the range of values observed in~\cite{Hegazy2015, 8473376}.
Note that all delays above \SI{50}{ms} (more than \SI{76}{percent} in this set)  will always be dropped.
Some results below \SI{50}{ms} will also be dropped due to the additional execution time.


The delay distribution in \Cref{fig:evaluation-distributions-plot}-A has a small effect on the controller.
Only a small difference is discernible between the red and green lines in \Cref{fig:evaluation-control-plot}-IV at around \SI{4}{s} and at the last set-point change.
The blue trajectory in \Cref{fig:evaluation-control-plot}-IV corresponds to the distribution in \Cref{fig:evaluation-distributions-plot}-B.
Clearly the latencies in this case have implications on the control.
The primary consequence is a delayed response to set-point changes.
As the set-point changes and all new \ac{MPC} requests are lost due to infeasibility or delay, the system keeps operating using open loop data from the previous set-point.
In this situation, results from a short horizon is preferable as otherwise reactions to new set-point changes will lag.
An edge strategy is used to improve the situation.
To keep the load on the edge down the system continuously sends the latest horizon used to the edge (the shortest feasible result received from the cloud in the previous iteration).
The shorter horizon also provides the highest probability of continuously receiving results to stay in closed loop mode when variable delay and execution time is considered for the edge.
If at any time the system does not receive new results and starts the process of switching to local mode, the next iteration instead places an optimization using the largest horizon at the edge.
For simplicity and illustration the edge is set to a fixed network delay of \SI{40}{ms} in the example.
This is a large delay and as can be concluded from \Cref{fig:evaluation-exectimes-plot} the edge will sometimes not respond when adding the execution time.
This, however, covers for most of the delayed response (orange line in \Cref{fig:evaluation-control-plot}-IV) while horizon selection over all is equal to that of no delay (II green and IV orange in \Cref{table:usage}).


\newcommand{\tikzcircle}[1]{\tikz[baseline=-0.8ex]\draw[#1,fill=#1,radius=2pt] (0,0) circle ;}
\npdecimalsign{.}
\nprounddigits{2}
\begin{table}[h]
	\centering
	\caption{Horizon selection for use cases in \Cref{fig:evaluation-control-plot}. The title row shows the percentage of samples in which the \ac{MPC} operate  in closed loop. The columns show the fraction of time in which input from a group of horizons is used,}	
	\begin{tabular}{l|l|l|l|l|l|l}%
		\bfseries Horizon & \bfseries II \tikzcircle{red} & \bfseries II \tikzcircle{green} & \bfseries III & \bfseries IV \tikzcircle{green}  & \bfseries IV \tikzcircle{blue} & \bfseries IV \tikzcircle{orange}\\
		& \nprounddigits{1}\begin{filecontents*}{data/evaluation/complex_cloud_short_mpconly/mpcinfo.csv}
99.002
\end{filecontents*}
\csvreader[no head]{data/evaluation/complex_cloud_short_mpconly/mpcinfo.csv}{1=\A}{\numprint{\A}}\%				
		& \nprounddigits{1}\begin{filecontents*}{data/evaluation/complex_cloud_long_mpconly/mpcinfo.csv}
98.504
\end{filecontents*}
\csvreader[no head]{data/evaluation/complex_cloud_long_mpconly/mpcinfo.csv}{1=\A}{\numprint{\A}}\%
		& \nprounddigits{1}\begin{filecontents*}{data/evaluation/complex_cloud_short_netdown/mpcinfo.csv}
80.549
\end{filecontents*}
\csvreader[no head]{data/evaluation/complex_cloud_short_netdown/mpcinfo.csv}{1=\A}{\numprint{\A}}\%				
		& \nprounddigits{1}\begin{filecontents*}{data/evaluation/lambdalike_upper/mpcinfo.csv}
97.007
\end{filecontents*}
\csvreader[no head]{data/evaluation/lambdalike_upper/mpcinfo.csv}{1=\A}{\numprint{\A}}\%		
		& \nprounddigits{1}\begin{filecontents*}{data/evaluation/complex_cloud_short_dropout_badnet/mpcinfo.csv}
75.561
\end{filecontents*}
\csvreader[no head]{data/evaluation/complex_cloud_short_dropout_badnet/mpcinfo.csv}{1=\A}{\numprint{\A}}\%
		& \nprounddigits{1}\begin{filecontents*}{data/evaluation/withedge/mpcinfo.csv}
96.509
\end{filecontents*}
\csvreader[no head]{data/evaluation/withedge/mpcinfo.csv}{1=\A}{\numprint{\A}}\%
		\\\hline
		\begin{filecontents*}{data/evaluation/mpctable.csv}
6,6,0.70823,0.69327,0,0.5985,0.1197,0.5586
7,9,0.082294,0.097257,0,0.067332,0.20948,0.17456
10,12,0.074814,0.069826,0,0.05985,0.2818,0.037406
13,15,0.064838,0.069825,0,0.1197,0.1596,0.099751
16,18,0.039901,0.049875,0,0.024938,0.089776,0.044887
19,22,0.01995,0.009975,0.99002,0.009975,0.12968,0.074813
\end{filecontents*}
\csvreader[no head, column count=8]{data/evaluation/mpctable.csv}{1=\A,2=\B,3=\C,4=\D,5=\E,6=\F,7=\G,8=\E}
		{\A-\B & \numprint{\C} & \numprint{\E} & \numprint{\F} & \numprint{\D} & \numprint{\G} & \numprint{\E}\\}
	\end{tabular}
	\label{table:usage}
\end{table}

\pgfplotsset{
	boxplot/hide outliers/.code={
		\def\pgfplotsplothandlerboxplot@outlier{}%
	}
}

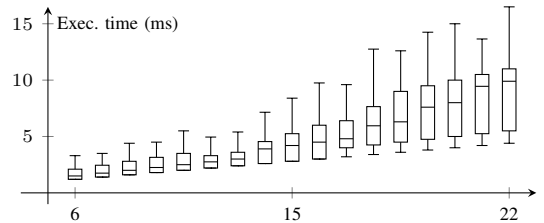
\begin{figure}[t]
    \centering
	\begin{tikzpicture}
	\begin{axis} [
	axis lines=center,
	xmin=-1,
	xmax = 18,
	ymin = -1,
	y=0.15cm,
	boxplot={draw direction=y, box extend=0.5,
		every box/.style = {solid,draw=black},
		every median/.style = {solid,draw=black},
		every whisker/.style = {solid,draw=black},
		hide outliers
	},
	ylabel={\scriptsize Exec. time (ms)},
	xtick={1,9,17},
	xticklabels={6,15,22},
	xticklabel style={font=\scriptsize},
	yticklabel style={font=\scriptsize}		
	]
	\addplot+ [boxplot] table[header=false, col sep=comma, y index=0] {data/evaluation/complex_cloud_short_dropout/exectimes.csv};
	\addplot+ [boxplot] table[header=false, col sep=comma, y index=1] {data/evaluation/complex_cloud_short_dropout/exectimes.csv};		
	\addplot+ [boxplot] table[header=false, col sep=comma, y index=2] {data/evaluation/complex_cloud_short_dropout/exectimes.csv};
	\addplot+ [boxplot] table[header=false, col sep=comma, y index=3] {data/evaluation/complex_cloud_short_dropout/exectimes.csv};	
	\addplot+ [boxplot] table[header=false, col sep=comma, y index=4] {data/evaluation/complex_cloud_short_dropout/exectimes.csv};
	\addplot+ [boxplot] table[header=false, col sep=comma, y index=5] {data/evaluation/complex_cloud_short_dropout/exectimes.csv};		
	\addplot+ [boxplot] table[header=false, col sep=comma, y index=6] {data/evaluation/complex_cloud_short_dropout/exectimes.csv};
	\addplot+ [boxplot] table[header=false, col sep=comma, y index=7] {data/evaluation/complex_cloud_short_dropout/exectimes.csv};			
	\addplot+ [boxplot] table[header=false, col sep=comma, y index=8] {data/evaluation/complex_cloud_short_dropout/exectimes.csv};
	\addplot+ [boxplot] table[header=false, col sep=comma, y index=9] {data/evaluation/complex_cloud_short_dropout/exectimes.csv};			
	\addplot+ [boxplot] table[header=false, col sep=comma, y index=10] {data/evaluation/complex_cloud_short_dropout/exectimes.csv};	
	\addplot+ [boxplot] table[header=false, col sep=comma, y index=11] {data/evaluation/complex_cloud_short_dropout/exectimes.csv};		
	\addplot+ [boxplot] table[header=false, col sep=comma, y index=12] {data/evaluation/complex_cloud_short_dropout/exectimes.csv};
	\addplot+ [boxplot] table[header=false, col sep=comma, y index=13] {data/evaluation/complex_cloud_short_dropout/exectimes.csv};		
	\addplot+ [boxplot] table[header=false, col sep=comma, y index=14] {data/evaluation/complex_cloud_short_dropout/exectimes.csv};	
	\addplot+ [boxplot] table[header=false, col sep=comma, y index=15] {data/evaluation/complex_cloud_short_dropout/exectimes.csv};		
	\addplot+ [boxplot] table[header=false, col sep=comma, y index=16] {data/evaluation/complex_cloud_short_dropout/exectimes.csv};
	\end{axis}
	\end{tikzpicture}
	\caption{Execution time distribution for the various control horizons in \Cref{fig:evaluation-control-plot}-IV. Outliers outside the $1.5\cdot$IQR ranges are excluded.}
	\label{fig:evaluation-exectimes-plot}
\end{figure}
\pgfplotsset{every axis title/.append style={at={(0.5,0.95)}}}
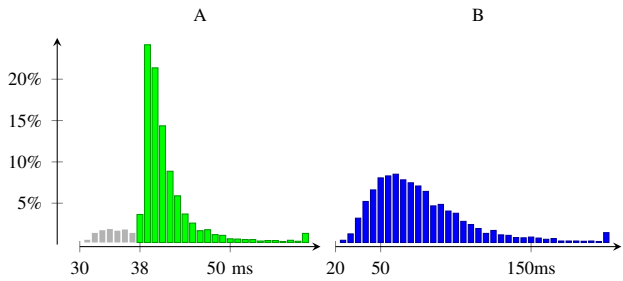
\begin{figure}[t]
    \centering
	\begin{tikzpicture}[trim left=-2mm]
	\begin{groupplot}[
		group style={group size=3 by 1,horizontal sep=0.2cm},
	]
	
		\nextgroupplot [ybar,
		bar shift=0,
		axis x line=none,
		axis y line=left,
		scaled ticks=false,
		ytick={0.05,0.1,0.15,0.2},
		yticklabels={5\%,10\%,15\%,20\%},
		yticklabel style={font=\scriptsize},
		ymax = 0.25,
		ymin = 0,
		xmax = 0,
		xmin = 0,
		bar width=1pt,		
		x=0.5mm,
		y=11cm
		]
	
	\nextgroupplot [ybar,
		bar shift=0,
		axis x line=bottom,
		axis y line=none,
		scaled ticks=false,
		ytick={0.05,0.1,0.15,0.2},
		yticklabels={5\%,10\%,15\%,20\%},
		xticklabel style={font=\scriptsize},
		yticklabel style={font=\scriptsize},
		xtick={30,38,50},
		xticklabels={30,38,50 ms},
		title={\scriptsize A},
		ymax = 0.25,
		ymin = -0.005,
		xmax = 62,
		xmin = 30,
		bar width=2.25pt,
		x=1mm,
		y=11cm
	]


	\addplot [restrict x to domain=38:60,draw=\goodnetcolor!60!black,fill=\goodnetcolor] table[header=false, col sep=comma, x index = 0, y index=1] {data/evaluation/lambda_like/net_delay_hist.csv};
	
	\addplot [restrict x to domain=0:37,draw=none,fill=black!30!white] table[header=false, col sep=comma, x index = 0, y index=1] {data/evaluation/lambda_like/net_delay_hist.csv};	
	
	\nextgroupplot [ybar,
		axis y line=none,	
		axis x line=bottom,
		scaled ticks=false,
		ytick=\empty,
		xticklabel style={font=\scriptsize},
		yticklabel style={font=\scriptsize},
		title={\scriptsize B},
		xtick={20,50,150},
		xticklabels={20,50,150ms},		
		ymax = 0.25,
		ymin = -0.005,
		bar width=2pt,
		x=0.2mm,
		y=11cm,
		xmin=20,
		xmax=210
	]

	\addplot [draw=\badnetcolor!60!black, fill=\badnetcolor] table[header=false, col sep=comma, x index = 0, y index=1] {data/evaluation/complex_cloud_short_dropout_badnet/net_delay_hist.csv};	
	
	\end{groupplot}[
	\end{tikzpicture}
	\caption{Latency distributions for the two network conditions in \Cref{fig:evaluation-control-plot}-IV. A) In green: Log-normal distribution with $\mu=0.8$, $\rho=0.8$ and an offset of \SI{38}{\ms} B) Log-normal distribution with $\mu=4$, $\rho=0.5$ and an offset of \SI{14}{\ms}}
	\label{fig:evaluation-distributions-plot}
\end{figure}
	\section{Conclusion}\label{sec:conclusions}
We have proposed and demonstrated a strategy for control systems operated using the cloud. The method aims to extend a local controller whenever cloud connectivity is present.
We show how this can be achieved using a combination of unconstrained and constrained control, in the form of \ac{LQR}, and \ac{MPC}.
We evaluated the approach on a simulated ball and beam process to show that an assisted controller can improve performance while being robust to connectivity issues.
The primary mode studied is one that in each sample issues requests for a range of prediction horizons and uses the results corresponding to the smallest feasible horizon returned in time.
This was combined with an edge strategy to reduce the impact of long delays.

The method allows for offloading complex control to the cloud and incorporating information not available locally.
It could also be used to enhance readily made designs and cover for unforeseen events. 
For instance, rate limiters and step size limiters could be augmented with a cloud assisted approach.
Another potential for an assisted cloud approach is that improvements can be incorporated into systems while keeping them operational.
Consider for instance an extension to the system model which increases the computational demand of the controller.
The new model is simply injected into the cloud \acp{MPC}.
Infeasible results or prolonged executions times will be discarded and not have a negative impact on the system.

A lot of interesting further work exists, e.g., 
empirical studies of extended evaluations and implementations,
improvement of the assisted approach in terms of a larger range of disturbances, tracking methods, and request/response selection methods.
There is also future work that relates to the presented approach, e.g., the variability in the definition of the \ac{MPC} requests, incorporating larger problem spaces, advanced implementations in the cloud back-end, and scheduling heuristics in the cloud.



	\ifunmasked
\section*{Acknowledgements}
This work is partially funded by the Wallenberg AI, Autonomous Systems and Software Program (WASP) funded by the Knut and Alice Wallenberg Foundation, SEC4FACTORY project SSF RIT17-0032 funded by the Swedish Foundation for Strategic Research (SSF), and the HI2OT University Network on Industrial IoT funded by Nordforsk. 
The authors are part of the Excellence Center at Linköping-Lund in Information Technology (ELLIIT).
\fi
	\bibliographystyle{IEEEtran}
	\bibliography{references}

\end{document}